\documentstyle[12pt]{article}
\begin{document}
\begin{center}
{\Large  On Pair--Particle Distribution\\[5mm]
in Imperfect Bose Gas}\\[5mm]
{\large A. A. Shanenko}\\[5mm]
{\it Bogoliubov Laboratory of Theoretical Physics\\[2mm]
Joint Institute for Nuclear Research\\[2mm]
Dubna, Russia}
\end{center}
\begin{abstract}
A simple model of estimating the radial distribution
function of an imperfect Bose gas in the ground state
is presented. The model is based on integro--differential
equations derived by considering the space boson distribution
in an external field. With the approach proposed,
the particular case of dilute Bose gas is investigated within
the hard sphere approximation and beyond.
\end{abstract}


\section{Introduction}

This letter is devoted to investigation of spatial particle
correlations in many--body systems in the spirit of the approach
of Ref.1 which consists in the following.
Let us consider a uniform many--body system with the
interaction potential $\Phi(r)$ at the particle density $n=N/V$.
As it has been shown for the classical case~\cite{shan}, the radial
distribution function of this system $g(r)$ can be calculated with
\begin{equation}
g(r)=\frac{n_{str}(r)}{n} \; ,
\label{eq1}\end{equation}
where $n_{str}(r)$ denotes the density of the mentioned particles
at the point $\vec r \;(r \equiv \mid \vec r \mid)$ of the nonuniform
structure appearing under the action of the external
field $\Phi(r)$ on the system considered. Here it is implied that
$$\Phi(r)\rightarrow 0\;,n_{str}(r) \rightarrow n\;
                             (r \rightarrow \infty)\;.$$
Relation~(\ref{eq1}) seems quite natural because the quantity
$g(\mid\vec r_1-\vec r_2\mid)/V$ is the probability of finding
the second particle of the system at the point $\vec r_2$ provided
that the first one is at the point $\vec r_1$. Therefore, we are
able to calculate $g(r)$ by considering the space distribution of the
particles with the numbers $2,...,N$ in the external field generated
by the first particle fixed at the origin of the coordinates.

The aim of the present article is to clarify to what extent
relation~(\ref{eq1}) is suitable for the case of an imperfect
Bose gas at zero temperature. The paper is organized as follows.
To simplify understanding of the subject, the results of
investigating relation(\ref{eq1}) in the classical case are
briefly discussed in the second section. The third paragraph
contains the main items of deriving integro--differential
equations for $g(r)$ of a Bose gas in the ground state on the
basis of (\ref{eq1}). To check validity of the equations, in the
fourth section the Bose gas of the hard spheres is considered
in the weak coupling approximation. At last, the fifth paragraph
is the consideration of a dilute Bose gas beyond the hard
sphere simplification.

\section{Classical Case}

To demonstrate reasonable character of relation~(\ref{eq1}) and
simplify understanding of further arguments, let us consider
a many--body system of classical particles interacting with each
other by means of the potential $\Phi(r)$ and being
in the external field $\Phi(r)$. In this case we are able to
calculate the structure function $n_{str}(r)$ with the well--known
Thomas--Fermi approximation~\cite{gomb}. According to this method,
$n_{str}(r)$ can be found with the following condition:
\begin{equation}
\mu\left( n_{str}(r) \right) + \Phi(r)
       + U(r)=const \;\; (\forall \vec r),
\label{eq2}\end{equation}
where $\mu(n_{str}(r))$ is the chemical potential of the ideal gas
of the given particles at the density $n_{str}(r)\;$; $\;U(r)$
denotes the energy of the interaction of the particle at the point
$\vec r$ with the surrounding particles. To complete the
calculational procedure, we use the reasonable integral relation
\begin{equation}
U(r)= \int_{V}\Phi(\mid \vec r-\vec y\mid)\;n_{str}(y) d\vec y\; .
\label{eq3}\end{equation}
Now, we have got everything to obtain an integral equation for
$g(r)$ in the classical case. At some point $\vec r_{far}$ being
far enough from the origin of the external field,
expression~(\ref{eq2}) is written as
\begin{equation}
\mu(n)+ n\int_{V}\Phi(\mid \vec r_{far}-\vec y\mid) d\vec y
= const
\label{eq4}\end{equation}
provided that we limit ourselves to the situation
$$\lim\limits_{r \to \infty} \Phi(r)=0\,.$$
Using formulae~(\ref{eq1})--(\ref{eq4}), the classical approximation
$$\mu(n) - \mu(n_{str}(r)) =
      - \theta \ln \left(n_{str}(r)/n\right) $$
and the relation
\begin{equation}
\lim\limits_{V\to \infty}\frac{\int_{V}
    \bigl(\Phi(\mid\vec r_{far}-\vec y\mid) -
        \Phi(\mid\vec r-\vec y\mid)\bigr) d\vec y}
           {\int_{V}\Phi(\mid\vec r-\vec y\mid) d \vec y}\;=\;0\;,
\label{eq5}\end{equation}
we arrive at
\begin{equation}
-\theta\, \ln g(r) =
        \Phi(r)+ n\int_{V}\bigl(g(y)-1\bigr)
               \Phi(\mid\vec r-\vec y\mid) d\vec y\;
\label{eq6}\end{equation}
Note that relation~(\ref{eq5}) is valid for lots of the known
potentials, in particular, for the integrable and
Coulomb---like potentials.

We have derived integral equation~(\ref{eq6}) neglecting
spatial particle correlations when calculating $U(r)$~\cite{shan}.
Therefore, it can only be used for the integrable and Coulomb--like
potentials but not for the strongly singular potentials with the
short range behaviour of the Lennard--Jones type. To include the
correlations, we should be based on a more accurate approximation
of $U(r)$. The simplest one is the following~\cite{shan}:
\begin{equation}
U(r)= \int_{V} g(\mid\vec r-\vec y\mid)
                   \; \Phi(\mid\vec r-\vec y\mid) \;
                                    n_{str}(y) d\vec y\; .
\label{eq7}\end{equation}
With~(\ref{eq1}),~(\ref{eq2}) and (\ref{eq7}) we find
\begin{eqnarray}
-\theta\,\ln g(r)=\Phi(r)+
n\int_{V}\bigl(g(y)-1\bigr) \;
          g(\mid \vec r-\vec y \mid) \;
                       \Phi(\mid \vec r-\vec y \mid)
                                         \;  d\vec y\; .
\label{eq8}\end{eqnarray}
In spite of the simplest version of approximating~$U(r)$ with the
space particle correlations taken into account, integral
equation~(\ref{eq8}) is very similar to the well--known Bogoliubov
equation for $g(r)$~(see discussion in Ref.1).
As it is seen, relations~(\ref{eq6}) and (\ref{eq8}) can be used
for the integrable, Coulomb--like as well as Lennard--Jones
potentials because $g(r)\Phi(r) \rightarrow 0\;(r \rightarrow 0)$
for the strongly singular $\Phi(r)$. Thus, in this paragraph we have
been convinced of the correctness of relation~(\ref{eq1}) for the
classical many--body systems. To elucidate the problem
of using~(\ref{eq1}) in the quantum case, we further
investigate many--boson system in the ground state.

\section{Bose Gas in the Ground State}

One could expect that in the quantum case being essentially
more complicated than the classical one, such a simple approach
based on expression~(\ref{eq1}) is not able to provide us with
satisfactory results. However, in the situation with no exchange
effects, for the Bose gas in the ground state, relation~(\ref{eq1})
yields a quite reasonable estimate of $g(r)$ with only one obvious
correction. To obtain more accurate data on $g(r)$ for a uniform
many--body system of bosons with the mass $m$, we need to
investigate a nonuniform Bose gas made of particles with the mass
$m/2\;.$ It should be noted, that as the quantity $m$ does
not appear in the integral equations of the previous paragraph,
the mentioned correction does not contradict the classical case
either.

Now, let us consider the system of $N-1$ zero-spin bosons with the
mass $m/2$ at zero temperature and in the external field $\Phi(r)$.
As before, we assume the interparticle potential to be $\Phi(r)$.
The simplest way to find $n_{str}(r)$ in this case is to adopt
the Hartree approximation for the wave function of the system
\begin{equation}
\psi_s=\prod\limits_{i=1}^{N-1} \;\psi(\vec r_i),
\label{eq9}\end{equation}
where $\psi(\vec r_i)$ denotes the normalized wave function of
the $i-$th boson in the ground state. As $\psi(\vec r)$ can be
chosen to be a real function~\cite{feyn}, we have
\begin{equation}
n_{str}(r)=(N-1)\,\psi^2(\vec r),
\label{eq10}\end{equation}
and, therefore, $\psi(\vec r)=\psi(r)$. Thus, to calculate
$n_{str}(r)$, we need evaluating $\psi(r)$ which obeys the
following Schr\"odinger equation:
\begin{eqnarray}
-\frac{\hbar^2}{m}\triangle\psi(r)
                  &+&\Phi(r)\psi(r)+ \nonumber\\ [3mm]
&&+(N-2)\psi(r)\int_{V}\Phi(\mid \vec r - \vec y \mid) \psi^2(y)
d\vec y = E_0 \psi(r).
\label{eq11}\end{eqnarray}
Here $E_0$ denotes the energy of a boson in the ground state. Not to
solve the eigenvalue problem, let us take into account that at
the point $\vec r_{far}$ being far enough from the origin of the
coordinates
\begin{equation}
\psi(r_{far}) = \frac{1}{\sqrt{V}}\;\;\;\;
                                    \bigl(\Phi(r_{far})=0\bigr).
\label{eq12}\end{equation}
Substituting (\ref{eq12}) into (\ref{eq11}) one can obtain
\begin{equation}
\frac{N-2}{V}\int_V\Phi(\mid \vec r_{far} - \vec y \mid)d\vec y
=E_0.
\label{eq13}\end{equation}
Further, with the use of relations~(\ref{eq5}), (\ref{eq11}) and
(\ref{eq13}) we arrive at
\begin{eqnarray}
\frac{\hbar^2}{m}\triangle\psi(r)&=&\Phi(r)\psi(r)+\nonumber\\[3mm]
&&+\frac{(N-2)}{V}\psi(r)\int_{V}
     \Phi(\mid \vec r - \vec y \mid)
               \left(V\psi^2(y)-1\right)d\vec y.
\label{eq14}\end{eqnarray}
Taking into account that in the thermodynamic limit
$$\frac{N-2}{V} \simeq \frac{N}{V}=n$$
and using (\ref{eq1}), (\ref{eq10}) and (\ref{eq14}),
we derive the following integro--differential equation for
$g(r)$ of a cold Bose gas:
\begin{eqnarray}
\frac{\hbar^2}{m g^{1/2}(r)}\triangle g^{1/2}(r)&=&\Phi(r)
                                           +\nonumber\\[3mm]
&&+n \int_{V}
     \Phi(\mid \vec r - \vec y \mid)
               \left(g(y)-1\right)d\vec y.
\label{eq15}\end{eqnarray}
By analogy with the classical case this equation can be generalized
to the situation of strongly singular potentials
\begin{eqnarray}
\frac{\hbar^2}{m g^{1/2}(r)}\triangle g^{1/2}(r)&=&\Phi(r)
                                           +\nonumber\\[3mm]
&&+n \int_{V}
       g(\mid \vec r - \vec y \mid)
            \Phi(\mid \vec r - \vec y \mid)
                    \left(g(y)-1\right)d\vec y.
\label{eq16}\end{eqnarray}
Remark that equations~(\ref{eq15}) and (\ref{eq16}) have been
derived for the spinless bosons but they are also valid for bosons
with nonzero spin.

\newpage
\section{Bose Gas with the Hard Sphere Interaction}

To verify the reasonable character of equations~(\ref{eq15})
and (\ref{eq16}), let us consider the well--known example of the
hard--sphere Bose gas in the weak coupling approximation
characterized by the relation
$$ g(r)=1-\varepsilon(r), \;\;\varepsilon(r) \ll 1.$$
Taking account of this relation and limiting oneself to the first
order in $\varepsilon(r)$, one can rewrite
equation~(\ref{eq15})~(or~(\ref{eq16})) as
\begin{equation}
-\frac{\hbar^2}{2m}\triangle \varepsilon(r)=\Phi(r)
-n \int_{V}
     \Phi(\mid \vec r - \vec y \mid)
               \varepsilon(y)d\vec y.
\label{eq17}\end{equation}
The new equation is solved with the Fourier transformation
and gives the following result:
\begin{equation}
g(r)=1-\frac{1}{(2\pi)^3} \int \;\frac{\widetilde{\Phi}
(q)\;\,\exp(\imath\,\vec q \, \vec r\,)}{\hbar^2 q^2/2m\;+\;n
\widetilde{\Phi}(q)}\; d\vec q \; ,
\label{eq18}\end{equation}
where $\widetilde{\Phi}(q)$ denotes the Fourier transform of
$\Phi(r)\; .$ Inserting $\displaystyle \widetilde{\Phi}(q)=
4 \pi \hbar^2 a/m$, where $a$ denotes the sphere radius, we
find
\begin{equation}
g(r)=1-\frac{4\,a}{r}\,f_{est}(x), \;\;x\equiv 4r\sqrt{\pi a n},
\label{eq19}\end{equation}
here $\displaystyle f_{est}(x)=0.5 \exp(-x/\sqrt{2})$. How well
this estimate is coordinated with the explicit expression
(see Ref.4)
\begin{equation}
g(r)=1-\frac{4\,a}{r}\,f_{expl}(x),
\label{eq20}\end{equation}
is shown in the following table:\\[3mm]
\begin{tabular}{|c|c|c||c|c|c||c|c|c|}
\hline
$x$ &$f_{est}(x)$&$f_{expl}(x)$&$x$&$f_{est}(x)$&$f_{expl}(x)$
&$x$&$f_{est}(x)$&$f_{expl}(x)$ \\
\hline
0.0&0.500&0.500&2.0&0.122&0.098&5.0&0.015&0.012\\
\hline
0.5&0.351&0.328&2.5&0.085&0.067&10&0.0004&0.001\\
\hline
1.0&0.246&0.217&3.0&0.060&0.046&20&$3.6\cdot10^{-6}$&0.0002\\
\hline
1.5&0.173&0.145&4.0&0.030&0.023& & & \\
\hline
\end{tabular}
\vspace{0.5cm}
As it is seen, estimate~(\ref{eq19}) is in reasonable agreement
with the explicit result~(\ref{eq20}) except the region of
large distances
$$r > \frac{10}{\sqrt{2}}\, r_c\, ,$$
where the correlation length
$$ r_c=\sqrt{\frac{1}{8\pi\,a\,n}}$$
is the same for both the (\ref{eq19}) and (\ref{eq20}) expressions.
Note that in the case of (\ref{eq19}) the correlation length
can be found from $\exp(-x/\sqrt{2})=\exp(-1)\,.$
Estimation~(\ref{eq19}) does not give the known
$1/r^4\,-$decay of $g(r)$ for $r \gg r_c$~\cite{isih,huang}.
Therefore, we need to be careful investigating the spectrum
of the low-lying excitations in a Bose gas with
$g(r)$ estimated from equations~(\ref{eq15}) and (\ref{eq16}).
However, the model proposed can yield satisfactory
evaluations for the main thermodynamic characteristics which
are essentially determined by the values of $g(r)$ at $r < r_c$
and $r \sim r_c$. For example, the ground state energy calculated
with~(\ref{eq19}) is given by
\begin{equation}
E=\frac{2\pi\,\hbar^2\,a}{m}\;\frac{N^2}{V}\;\left(1+4.01
\sqrt{\frac{a^3\,N}{V}}\right)\;.
\label{eq21}\end{equation}
The explicit result~(\ref{eq20}) leads~\cite{isih1} to the expression
\begin{equation}
E=\frac{2\pi\,\hbar^2\,a}{m}\;\frac{N^2}{V}\;\left(1+4.81
\sqrt{\frac{a^3\,N}{V}}\right)\;
\label{eq22}\end{equation}
first found by Lee and Yang~\cite{lee}. Besides, we can improve
our estimates treating the boson mass in equations~(\ref{eq15}) and
(\ref{eq16}) as a free parameter of the model. In particular, to
derive~(\ref{eq22}) we should replace $m$ by $1.13m$ in~(\ref{eq15})
or (\ref{eq16}).

\section{Natural Dilute Bose Gas}

A helpful feature of equations~(\ref{eq15}) and (\ref{eq16}) consists
in the possibility of operating with them beyond the weak coupling 
and hard spheres. In particular, in investigating a natural dilute 
Bose gas the expansion in powers of $n$ for $g(r)$ may be of 
interest. To find it let us assume that
\begin{equation}
g^{1/2}(r)=u_0(r) + n\,u_1(r) + \ldots
\label{eq23}\end{equation}
and substitute~(\ref{eq23}) into, say, equation~(\ref{eq16}). We
arrive at the following differential equations:
\begin{equation}
\frac{\hbar^2}{m}\triangle u_0(r) - \Phi(r)\,u_0(r)=0,
\label{eq24}\end{equation}
and
\begin{eqnarray}
\frac{\hbar^2}{m}\triangle u_1(r)&-&
              \Phi(r)\,u_1(r)=\nonumber\\
&&=u_0(r)\int_V\,u^2_0(\mid \vec r - \vec y \mid)\,
	  \Phi(\mid \vec r - \vec y \mid)\,
	       (u^2_0(y) - 1) d\vec y.
\label{eq25}\end{eqnarray}
Relation~(24) is exactly the equation which is often used to find
the zero density limit for the function of the boson two--body
distribution~\cite{bogol}. Equations~(\ref{eq24}) and (\ref{eq25})
testify to that $g(r)$ of a natural Bose gas in the
ground state is an analytical function of $n$ contrary to the
case of the weakly interacting Bose gas of the hard spheres.

\section{Summary}

The integro--differential equations for the radial distribution
function of a cold Bose gas with the interaction potential
$\Phi(r)$, have been found via investigating the spatial
boson distribution in the external field $\Phi(r)\, .$
Checking the validity of these equations, we have considered
the weakly interacting Bose gas of the hard spheres and derived
the reasonable picture of the two--body correlations and
the satisfactory estimate of the mean energy. In the case of a
natural Bose gas~(~beyond the hard spheres and weak coupling~),
the equations have provided us with the expansion in powers of $n$
for $g(r)$ that can be useful at investigating a dilute
many--boson system.
\\[5mm]
{\it Acknowledgement:} The author thanks A. Yu. Cherny for helpful
and stimulating discussions.

\end{document}